\documentclass[conf]{new-aiaa}
\usepackage[utf8]{inputenc}

\usepackage{graphicx}
\usepackage{amsmath}
\usepackage[version=4]{mhchem}
\usepackage{siunitx}
\usepackage{longtable,tabularx}
\usepackage{subcaption}
\usepackage{hyperref}
\usepackage{textcomp}
\usepackage{multirow}

\title{Towards Distributed Semi-speculative Adaptive Anisotropic Parallel Mesh Generation}

\author{Kevin Garner\footnote{Research Assistant}, Christos Tsolakis\footnote{Research Assistant}, Polykarpos Thomadakis\footnote{Research Assistant} and Nikos Chrisochoides\footnote{Richard T. Cheng Chair Professor of Computer Science, email: nikos@cs.odu.edu}}
\affil{Center for Real-time Computing, Old Dominion University, Norfolk, VA 23529, USA}

\begin{document}

\maketitle

\begin{abstract}
This paper presents the foundational elements of a distributed memory method for mesh generation that is designed to leverage concurrency offered by large-scale computing. To achieve this goal, meshing functionality is separated from performance aspects by utilizing a separate entity for each - a shared memory mesh generation code called CDT3D and PREMA for parallel runtime support. Although CDT3D is designed for scalability, lessons are presented regarding additional measures that were taken to enable the code's integration into the distributed memory method as a black box. In the presented method, an initial mesh is data decomposed and subdomains are distributed amongst the nodes of a high-performance computing (HPC) cluster. Meshing operations within CDT3D utilize a speculative execution model, enabling the strict adaptation of subdomains' interior elements. Interface elements undergo several iterations of shifting so that they are adapted when their data dependencies are resolved. PREMA aids in this endeavor by providing asynchronous message passing between encapsulations of data, work load balancing, and migration capabilities all within a globally addressable namespace. PREMA also assists in establishing data dependencies between subdomains, thus enabling "neighborhoods" of subdomains to work independently of each other in performing interface shifts and adaptation. Preliminary results show that the presented method is able to produce meshes of comparable quality to those generated by the original shared memory CDT3D code. Given the costly overhead of collective communication seen by existing state-of-the-art software, relative communication performance of the presented distributed memory method also shows that its emphasis on avoiding global synchronization presents a potentially viable solution in achieving scalability when targeting large configurations of cores.
\end{abstract}

\section{Nomenclature}

{\renewcommand\arraystretch{1.0}
\noindent\begin{longtable*}{@{}l @{\quad=\quad} l@{}}
$S$ & subdomain of a partitioned grid\\
$I_{1,2}$ & interface boundary between subdomains 1 and 2 \\
$\begin{mathcal}M\end{mathcal}$  & continuous metric field \\
$M$ & discrete metric field defined at the vertices of a grid \\
$C(\cdot)$& complexity of a metric field \\
$L_a$ & Euclidean edge length evaluated in the metric of vertex \textit{a} \\
$M_{mean}$ & metric tensor interpolated at the centroid of a tetrahedron \\
$|k|$ & volume of a tetrahedron in evaluated metric $M_{mean}$ \\
$Q_k$ & mean ratio shape measure \\
\end{longtable*}}

\section{Introduction}
This paper presents the foundational elements of a distributed memory method for mesh generation that is designed to leverage concurrency offered by large-scale computing. To achieve this goal, meshing functionality is separated from performance (scalability) aspects by utilizing a separate entity for each - CDT3D \cite{Drakopoulos19F, TsolakisCDT3D} for mesh generation and PREMA \cite{BarkerPREMA, Thomadakis18PREMA, Thomadakis22PREMA} for parallel runtime support. CDT3D is a shared memory code that is intended to be used as a "black box" for scalability up to thousands of cores. Its meshing operations are designed to operate within the broader scope of scalable data-parallel and partially coupled methods within a framework we term the Telescopic Approach \cite{Chrisochoides2016TelescopicAF}. The Telescopic Approach provides a layout of multiple memory hierarchies within an exascale architecture and how different meshing kernels might be utilized at each level to achieve maximum concurrency. CDT3D is specifically built to operate within the lowest level of the hierarchy, exploiting fine-grain parallelism at the core and node levels. Although designed for scalability, lessons are presented regarding design oversight given that this code is optimized for execution within a single multicore node, and what additional measures were taken to enable the code’s integration into the distributed memory method as a black box. Whereas CDT3D targets the chip level, the distributed memory method serves to exploit coarse-grain parallelism at the node level. 

In the presented method, an initial mesh is data decomposed and subdomains are distributed amongst the nodes of a high performance computing (HPC) cluster. Meshing operations within the shared memory code are designed to adopt a speculative execution model, enabling the strict adaptation of interior subdomain elements so that interface elements can be adapted in a separate step to maintain mesh conformity. Interface elements undergo several iterations of shifting so that they are adapted when their data dependencies are resolved. To aid in this endeavor, a runtime system called PREMA is utilized, which alleviates the burden of work scheduling and load balancing for distributed memory applications. This system provides constructs which enable asynchronous message passing between encapsulations of data, work load balancing, and migration capabilities all within a globally addressable namespace. PREMA offers a particularly useful "event" feature which aids in establishing data dependencies between subdomains, thus enabling "neighborhoods" of subdomains to work independently of each other in performing interface shifts and adaptation. The distributed memory CDT3D method is designed to avoid the use of collective communication techniques that are utilized in existing state-of-the-art methods \cite{LOSEILLEFefloa, ParkRefine} due to the fact that they have been shown to hinder potential scalability \cite{ECPMPIStudy, Klenk2017ECPStudy, TsolakisEvaluation}. Preliminary results show that after several passes of interface shifts and adaptation, the presented method is able to produce meshes of comparable quality to those generated by the original shared memory CDT3D code. Relative communication performance also suggest that the interface shift operation presents a potentially viable solution in achieving scalability for mesh adaptation when targeting configurations with large numbers of cores (building upon the shared memory CDT3D method which only utilized up to 40 cores in an earlier study \cite{TsolakisEvaluation}).

\section{Background}
\subsection{Shared Memory Mesh Generation}
The shared-memory CDT3D implements a tightly-coupled method and exploits fine-grain parallelism at the cavity level using data decomposition, targeting shared memory multicore nodes using multithreaded execution at the chip level. Multiple mesh operations (e.g., point insertion, edge/face swapping, etc.) are performed concurrently on different data by using fast atomic lock instructions to guarantee correctness. These locks are used to acquire the necessary dependencies for the corresponding operation. Failure to do so will result in unlocking any acquired resources (rollback) and attempting to apply an operator on a different set of data. This is the essence of the speculative execution model, which is to exploit parallelism “everywhere possible” from the beginning of refinement when there is no, or very coarse, tessellation (contrary to existing methods that require sequential preprocessing and are in some cases just as expensive as the parallel mesh refinement itself). The speculative execution model is implemented using the separation of concerns ideology \cite{DijkstraSeparation} \cite{Tsolakis21TaskingFramework}. As mentioned previously, functionality is separated from performance components, even at the lowest level of the Telescopic Approach with CDT3D as well. Meshing operations are abstracted as tasks, and these tasks are only performed when their corresponding dependencies are satisfied (i.e. successfully locked). Such abstractions provide easy interoperability with a low level runtime system such as PREMA (discussed in more detail in section \ref{parallel_runtime_system}).

\begin{figure}[htb]
\begin{center}
\includegraphics[width=0.65\textwidth]{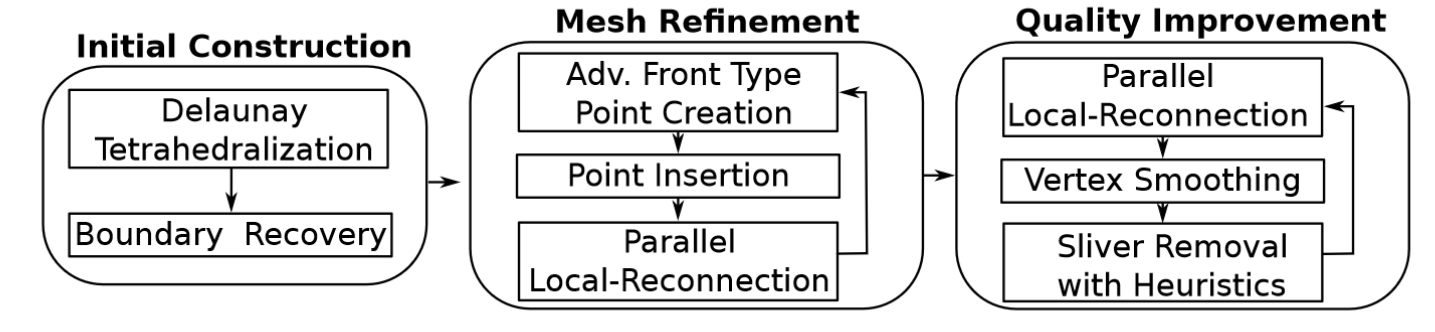}
\end{center}
\caption{CDT3D Pipeline of Isotropic Grid Generation \cite{Drakopoulos19F}}
\label{fig:cdt3d-isotropic-pipeline}
\end{figure}

One capability of CDT3D is to generate boundary-conforming isotropic tetrahedral meshes with element sizes defined by a point distribution function. The pipeline for isotropic mesh generation can be divided into three steps (also seen in Figure \ref{fig:cdt3d-isotropic-pipeline}): initial grid construction, grid refinement, and grid quality improvement. CDT3D was compared with AFLR in terms of its qualitative and quantitative results in its isotropic grid generation \cite{Drakopoulos19F}. Across several aerospace configurations, CDT3D exhibited comparable quality to all those generated by AFLR.

\begin{figure}[htb]
\begin{center}
\includegraphics[width=0.45\textwidth]{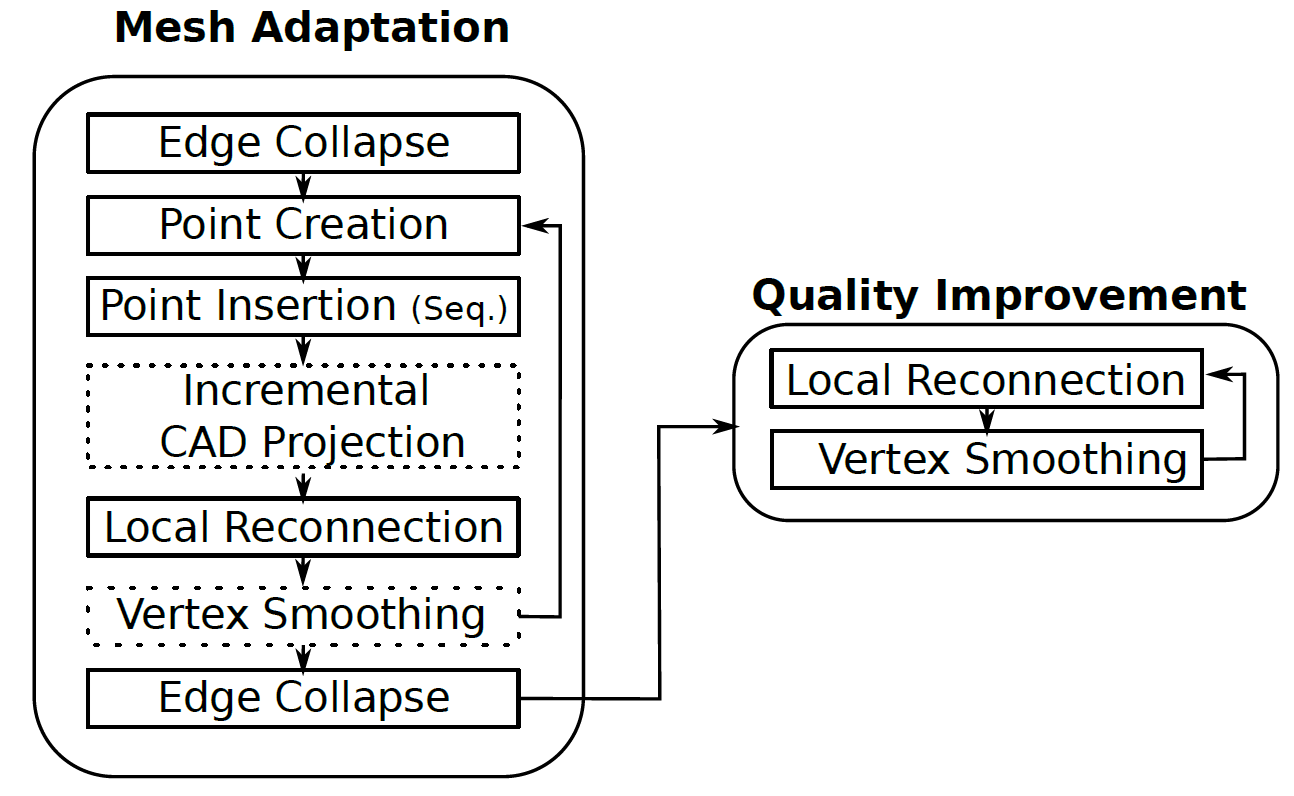}
\end{center}
\caption{CDT3D Pipeline of Anisotropic Adaptive Grid Generation \cite{TsolakisDissertation}}
\label{fig:cdt3d-anisotropic-pipeline}
\end{figure}

CDT3D also offers metric-based anisotropic mesh adaptation, where the metric can be derived from analytic or discrete fields, and can be combined with Computer-Aided Design (CAD)-based information to accomplish adaptation \cite{TsolakisCDT3D}. This pipeline can be seen in Figure \ref{fig:cdt3d-anisotropic-pipeline}.

CDT3D was compared with three other parallel anisotropic mesh adaptation methods used extensively within the industry \cite{TsolakisEvaluation}. Again, the quantitative and qualitative results of each method were compared from testing on a benchmark created by the Unstructured Grid Adaptation Working Group (UGAWG) \cite{UGAWG}. This benchmark served to evaluate adaptive mesh mechanics for analytic metric fields on planar and simple curved domains. In each case, CDT3D was shown to maintain stability of metric conformity (with both element shape size and edge length). It also showcased good performance when utilizing up to 40 cores on a single multicore node, exhibiting good weak scaling speedup and almost linear speedup amongst its strong scaling cases.

\subsection{Parallel Runtime System} \label{parallel_runtime_system}

Message passing and data migration within the presented distributed memory method is handled by utilizing the Parallel Runtime Environment for Multicore Applications (PREMA) system \cite{BarkerPREMA, Thomadakis18PREMA, Thomadakis22PREMA}. This system provides work scheduling and load balancing on both shared and distributed memory architectures, alleviating the application developer of these responsibilities. PREMA introduces constructs called mobile objects, which are encapsulations of data (not necessarily residing in contiguous memory), and mobile pointers which are used to identify mobile objects within a global namespace. Interactions between data can be expressed as remote method invocations (handlers) between mobile objects rather than between processes or threads. While non-conflicting handlers are executed concurrently across processing elements, PREMA offers the ability to utilize multiple hardware threads to share work within the context of individual handlers \cite{Thomadakis23TowardRuntime}. Due to the nature of adaptive applications, especially in the context of mesh generation, workload disparity is often witnessed amongst the mobile objects processed by handlers. This fine-grain parallelism allows for more efficient utilization of shared memory resources to help bridge this disparity in workload processing time. 

Distributed memory load balancing is achieved by PREMA’s ability to monitor work loads between ranks and perform migrations of mobile objects to available workers without interrupting execution. Communication and execution are separated into different threads to provide asynchronous message reception and instant computation execution at the arrival of new work requests. As stated previously, the mobile pointer construct allows method invocations to be made to mobile objects regardless of their location (potential migration to another rank). PREMA also provides the ability to establish dependencies between mobile objects and to execute user-defined events once all dependencies have been satisfied. This functionality becomes particularly useful when migrating cavity data needed for the adaptation of interface boundary elements within subdomains (discussed in more detail in section \ref{high_level_algorithm}).

\section{Related Work} \label{related_work}
The presented distributed memory method is motivated by past investigations of “black-box approaches," where parallelization was attempted for mesh generation programs (most of them were originally sequential) while making the least amount of modifications possible to their source code. One such approach is termed "functionality-first," which involves the parallelization of state-of-the-art mesh generation software that are fully functional and optimized for single-core architectures. Several studies addressing the viability of functionality-first black box approaches included VGRID \cite{ZagarisVGRID}, TetGen \cite{Chrisochoides18PDR}, and AFLR \cite{GarnerThesis}. Another black-box approach focused on the integration of a shared memory method, called PODM \cite{PODMFOTEINOS20142}, into a distributed memory framework \cite{PODMExperience}. PODM is a Delaunay-based mesh generation method that was not originally designed for execution within a distributed setting (as opposed to the presented method which utilizes local reconnection techniques and is designed to fit the Telescopic Approach). This approach strictly utilized PODM as a black box and involved the migration of large amounts of data (entire subdomains) throughout execution in order to resolve data dependencies that were required to satisfy the Delaunay property. This overhead accounted for more than 50\% of execution time, making this distributed approach 7x slower than the shared memory PODM when utilizing the same numbers of cores \cite{PODMExperience}. Evaluations of these black-box approaches exhibited the recurrent conclusion that if a code is not originally designed for scalability, it cannot be simply integrated into a parallel framework as a black box. Rather than devoting significant amounts of time to redesigning such codes, these studies encourage the development of "scalability-first" approaches (those designed with scalability as the focus and functionality added as needed), if one wishes to leverage the maximum potential speedup offered by large-scale architectures. 

A goal of the presented distributed memory method is to avoid collective communication, as this is not ideal for large-scale computing. The overhead of collective communication was reported in an extensive study involving numerous proxy applications within DoE's Exascale Computing Project (ECP) \cite{ECPMPIStudy, Klenk2017ECPStudy}. The purpose of the ECP study is to understand communication patterns utilized by these applications and to identify where optimization efforts may be focused. It was observed that most of the applications spent more than 50\% of their runtime in communication (as opposed to computation). While the majority of communication calls were primarily point-to-point (messages between individual processes), the amount of runtime spent in communication was dominated by collective calls (e.g. MPI\_Allreduce, MPI\_Alltoall, etc.). 

Another observation of note is that none of the applications in the study utilized neighborhood collectives, a feature introduced in MPI 3.0 that permits collective communication calls within subgroups of processing elements \cite{MPINeighborCollectives}. This feature is designed to operate based on a process topology. Neighborhood collectives, in addition to several other MPI functions, are suggested as potential resources in helping drive forth optimizations for these exascale applications represented by their proxy counterparts.

The parallel meshing strategies utilized by the aforementioned state-of-the-art codes (that were compared to the shared memory CDT3D) all exhibit good speedup in the strong scaling and weak scaling cases presented in \cite{TsolakisEvaluation}. However, there are implicit global synchronization points in these codes that induce increasingly noticeable overhead when utilizing up to several hundred cores. Given the observations presented in the study of DoE's ECP, this overhead may be exacerbated when meshing billions of elements on much higher configurations of cores. For example, \textit{Feflo.a}, a functionality-first, partially-coupled coarse-grained approach developed by Inria \cite{LOSEILLEFefloa}, solves the interface problem by freezing interfaces during adaptation and then re-partitioning the domain to focus on adapting interface elements. This re-partitioning occurs only after all subdomains have completed adaptation and is repeated over several passes. This domain decomposition was reported to be one of the main parallel overheads. A scalability-first, partially-coupled coarse-grained approach developed by NASA, called \textit{refine} \cite{ParkRefine}, also re-partitions the domain at the end of each adaptation pass. Interior elements are adapted while interfaces initially remain frozen. Then, the method focuses on adapting interface elements while simultaneously performing communication to update neighboring subdomains of the changes to shared interface elements. Global identifiers are utilized to denote duplicate points between subdomains. Moreover, all-to-all communication occurs at the end of an adaptation pass. Each subdomain communicates with all other subdomains to ensure that each newly inserted grid point has a unique global identifier.

The distributed memory method presented aims to avoid collective communication, and does not perform global re-partitioning. Additionally, it does not require global synchronization to update global identifiers for duplicate data between subdomains. It instead takes a similar approach to \textit{EPIC} \cite{EPIC2012}, a functionality-first, partially-coupled coarse-grained approach developed by Boeing. \textit{EPIC} freezes interfaces during an initial adaptation pass and then shifts the interface elements into the interiors of subdomains. Only a subset of mesh operations within \textit{EPIC} utilize multi-threading however and the software does not take a speculative execution approach, similar to the shared memory CDT3D software, for individual subdomains. The presented distributed memory method performs an adaptation pass with frozen interfaces (fully utilizing the multithreaded speculative execution model for adapting interior elements within each subdomain), and then shifts cavities of data needed to adapt those interface elements between subdomains over several passes. Message passing is performed within neighborhoods of subdomains, avoiding any all-to-all communication.

\section{Distributed Memory Method}
\subsection{High-level Algorithm} \label{high_level_algorithm}
\begin{figure}[htb]
\begin{center}
\includegraphics[width=0.6\textwidth]{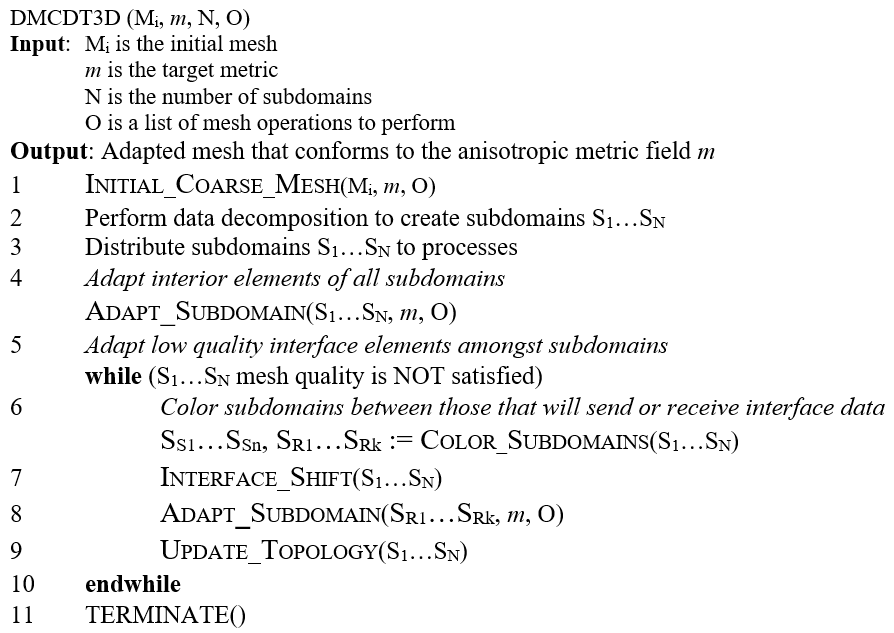}
\end{center}
\caption{High-level Algorithm of Early Distributed Memory CDT3D Implementation}
\label{fig:high-level-algorithm}
\end{figure}

Given its scalable design and performance in stability, the shared memory CDT3D (SMCDT3D) was abstracted as a library to be used in the adaptation of individual subdomains in the distributed memory method. A high-level overview of the distributed memory method (DMCDT3D) is shown in Figure \ref{fig:high-level-algorithm}. It essentially includes six steps, the latter three of which are executed in a loop until the quality of the entire mesh is satisfied. These steps include: initial coarse mesh generation (which is optional if a geometry volume is provided as input), decomposition, interior refinement/adaptation of all subdomains, interface shift, interior refinement/adaptation of colored subdomains, and a topology update of subdomain adjacency.

Different methods of decomposition may be applied to the grid; moreover, the method of decomposition applied in the current implementation is called PQR, which uses a sorting-based method to partition elements into subdomains based on a boundary-conforming curvilinear coordinate system \cite{CHRISOCHOIDES199475}. A geometry graph partitioning heuristic is utilized, where the mesh itself is considered to be a Euclidean graph (the elements are vertices and their face-connected neighbors establish edges). This heuristic uses Euclidean metrics and minimizes the diameter of subdomains, delivering quasi-uniform partitions. Once all subdomains have been created, they are packed and distributed among processes using PREMA.

After data decomposition, the interior elements within each subdomain are adapted. Whenever a subdomain undergoes adaptation, all of its interface elements are frozen. Let a mesh be partitioned into N subdomains \(S_1...S_N\). If \(S_1 \cap S_2 \neq \emptyset\) then \(S_1\) and \(S_2\) are neighbors. Let \(I_{1,2} = S_1 \cap S_2\). All points in \(I_{1,2}\) are interface points. Any edge, face, or tetrahedron defined by an interface point in \(I_{1,2}\) is an interface edge, interface face, or interface tetrahedron, respectively. Any other subdomain which contains elements that are defined by a point in \(I_{1,2}\) is also considered a neighbor of both \(S_1\) and \(S_2\) and will contain a copy of that same interface point.

Let \(neighs\) be the set of all neighbor subdomains for \(S_1\). The set of interior points in \(S_1\) is defined as: \(\forall points \in S_1 and \notin I_{1,neighs}\). Elements that are solely defined by interior points are permitted to undergo adaptation. Interface elements remain frozen to maintain conformity between subdomains. 

\begin{figure}[htb]
     \centering
     \begin{subfigure}[htb]{0.3\textwidth}
         \centering
         \includegraphics[width=\textwidth]{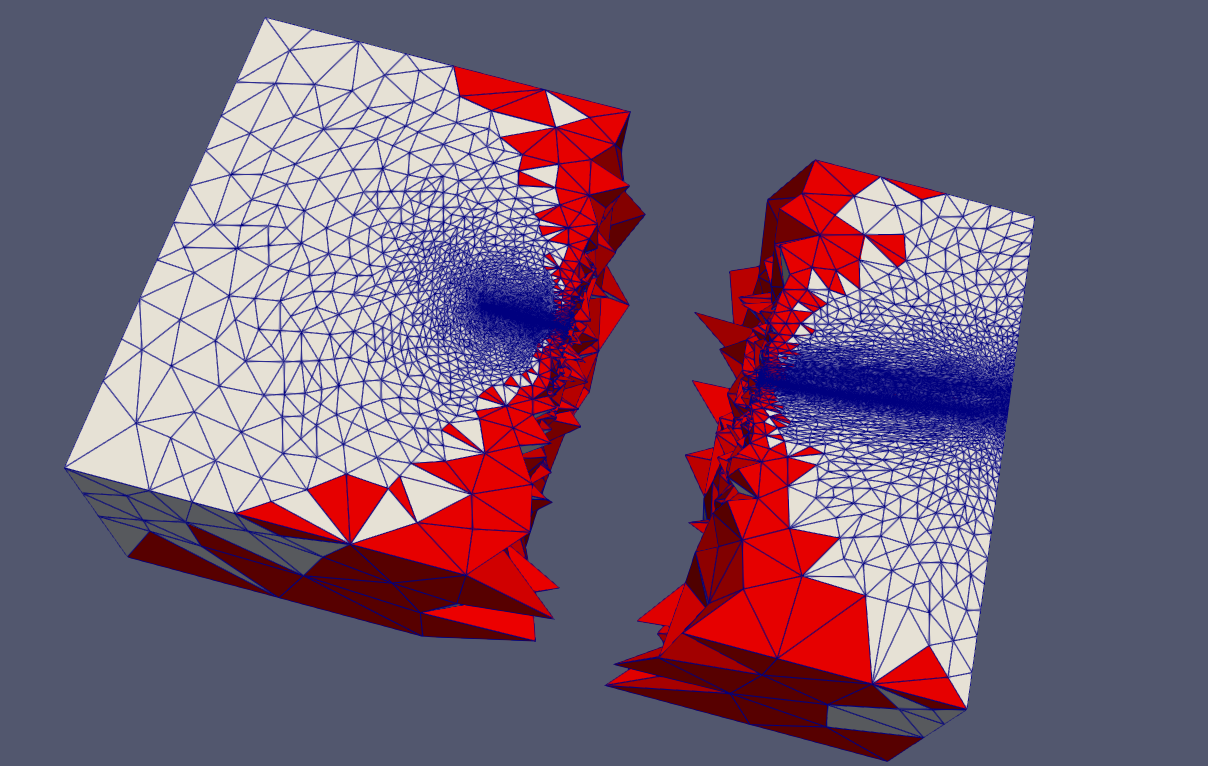}
         \caption{Data Decomposition}
         \label{fig:data_decomposition}
     \end{subfigure}
     \hfill
     \begin{subfigure}[htb]{0.3\textwidth}
         \centering
         \includegraphics[width=\textwidth]{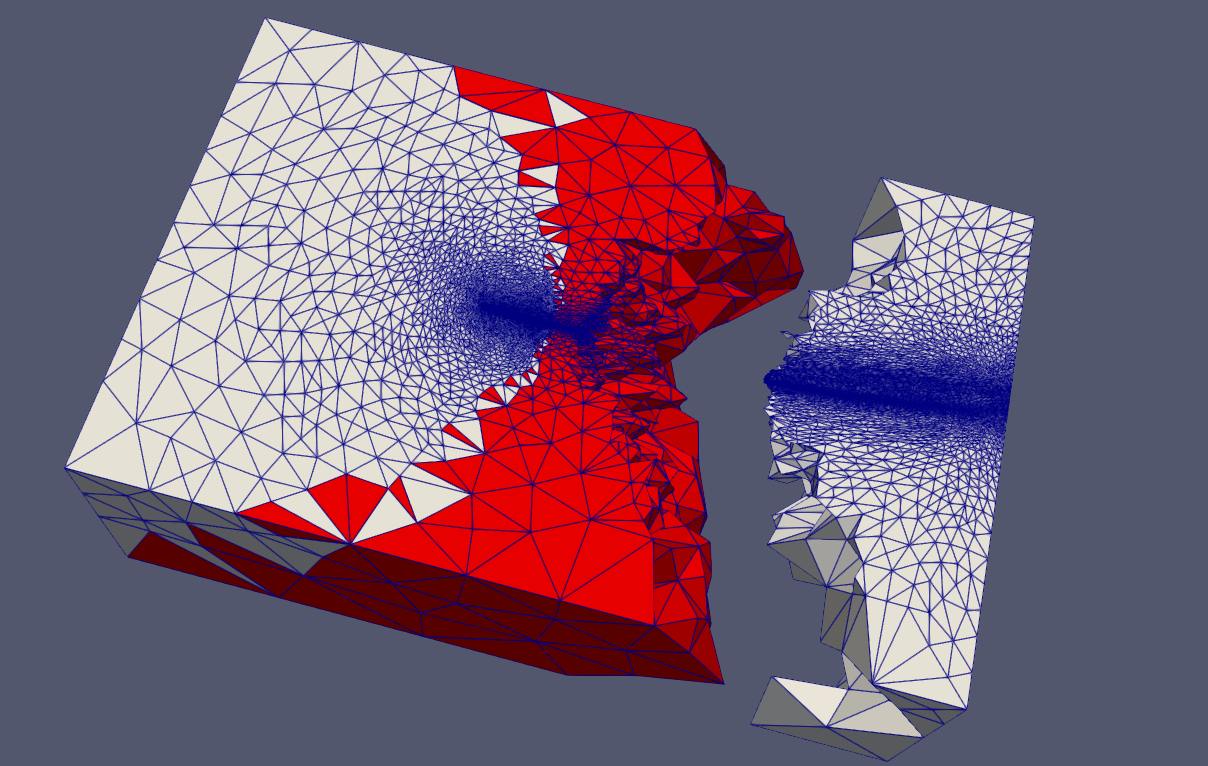}
         \caption{Interface Shift}
         \label{fig:interface_shift}
     \end{subfigure}
     \hfill
     \begin{subfigure}[htb]{0.3\textwidth}
         \centering
         \includegraphics[width=\textwidth]{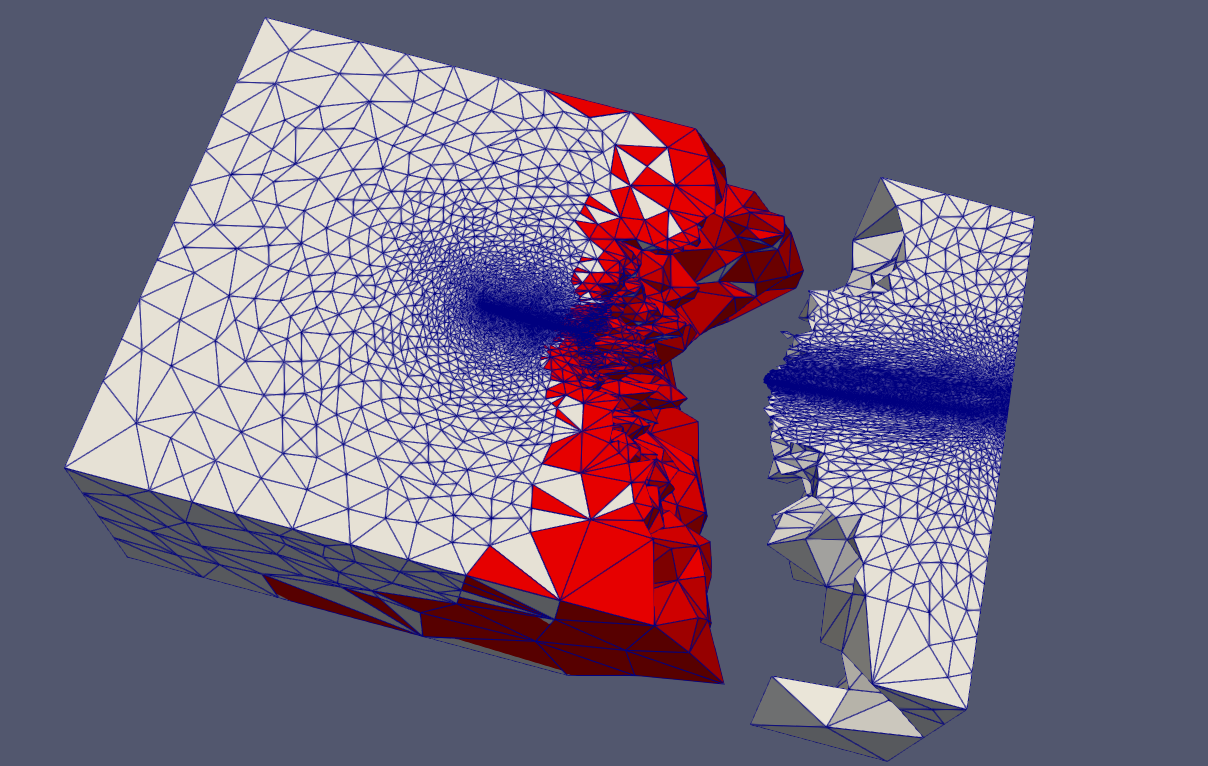}
         \caption{Mixed Interior/Interface Adaptation}
         \label{fig:interface_refinement}
     \end{subfigure}
        \caption{The interface shift and adaptation process is shown on a data decomposition of a delta wing geometry. Red elements contain edges that do not conform to the target metric, as they are affected by the frozen interface. (a) shows a data decomposition where two subdomains have each undergone interior adaptation. (b) shows the transfer of interface elements (and corresponding cavities) from the right subdomain to the left subdomain. (c) shows the adaptation of those elements that were previously interface elements in the left subdomain.}
        \label{fig:interface_shift_refinement_process}
\end{figure}

Due to how interface elements can affect the overall quality of the final mesh (as seen in the study regarding the parallelization of AFLR) \cite{GarnerThesis}, they are shifted to the interiors of subdomains over several iterations so that they may undergo adaptation. This mixed interior/interface (MII) adaptation phase begins immediately after a subdomain has completed its initial stage of interior adaptation. Subdomain adjacency can be considered as an undirected graph, where a subdomain is a vertex and an edge is a neighbor connection between two subdomains that share an interface point. Many subdomains within this graph are selected, or "colored," to receive data while particular neighbors are selected to send their corresponding interface data to those that are colored. Subdomains colored to receive data are prioritized by their number of low quality elements. The interface data sent consists not only of the interface elements themselves but also their corresponding cavities plus several layers of elements that are required to successfully permit their adaptation in each meshing operation. Consider a single tetrahedron as a single layer. The set containing each tetrahedron connected to each of its faces are considered to be a second layer, and all of the tetrahedra connected to each of their faces are considered to be the third layer, and so on. PREMA’s aforementioned event feature is utilized, where colored subdomains designate their neighbors as dependencies. Once this event activates (neighbor subdomains have completed their own interior adaptation), all neighbor subdomains gather interface data, send it to their colored counterpart, and the receiver scatters the interface data. Once all data has been received and scattered, the colored subdomain initiates interior adaptation. Those elements that were considered interfaces in the previous iteration are now considered to be interior elements due to this "padding" of cavities and layers of additional elements. These elements combined with the original interior elements will all now be processed for adaptation. Figure \ref{fig:interface_shift_refinement_process} shows an example of this process with a data decomposition of a delta wing geometry.

As in the shared memory method (which requires multiple grid generation passes of adaptation to obtain satisfactory grid quality, i.e. metric conformity) \cite{Drakopoulos19F, TsolakisCDT3D}, several iterations of interface shift and adaptation passes are required to attain sufficient grid quality in the distributed memory method. Before an iteration of MII adaptation can commence for a set of subdomains, the graph of subdomain adjacency must be up-to-date to ensure correct communication between subdomains. After a gather/scatter operation, subdomain adjacency may change (i.e., loss of adjacency or new adjacency). Each subdomain must communicate with its prospective neighbors to verify which of them are still neighbors, which are new neighbors (if any), and which are no longer neighbors. The distributed memory method performs this communication between subdomains asynchronously while other subdomains may still be undergoing adaptation. This subdomain adjacency update algorithm, and how it guarantees up-to-date adjacency before subsequent interface shift iterations, will be shown in the finalized manuscript of this work. Once the quality of all elements in each subdomain (interior plus interface) are found to be satisfactory, interface shifting ends, the final mesh is output, and the distributed memory method terminates.

Additional challenges regarding the interface shift and MII adaptation include: (1) determining how many interface shift iterations are required to attain satisfactory quality when subdomains are colored to send data based on interface elements that are point-connected to the receiving subdomains vs. gathering data using only elements that are face-connected and (2) how to maintain simple connectivity within subdomains (i.e. ensuring that no tetrahedron, or partition of tetrahedra, are connected to others by only a point or an edge within a subdomain) given that the shared memory software is designed to process only manifold geometries. Gathering data based on face-connectivity helps to ensure simple connectivity (although it doesn't guarantee it); however, gathering based on point-connectivity ensures that all mesh operations will have the needed cavities for interface elements (some operations, such as vertex smoothing, operate on points and edge connections rather than tetrahedra and face connections). Gathering based on point-connectivity collects more elements, giving the shared memory code more flexibility to achieve adaptation, allowing the distributed memory method to generate a mesh of sufficient quality quicker (as opposed to performing more interface shifts and phases of MII adaptation). The challenge of performing point-connected gathering while maintaining simple connectivity amongst all subdomains has been addressed in the current implementation and will be further detailed in the finalized manuscript of this work.

\subsection{Distributed Data Structures}

One should identify and encapsulate only data that is essential for jumpstarting the shared memory code in a distributed setting. There is a clean separation between the shared memory code and distributed code. The shared memory code itself is not distributed-aware. It is rather utilized as a library, where a number of adaptation operations can be executed on the interior of each subdomain while keeping interface elements fixed. Therefore, information pertaining to mesh elements within each subdomain must be extracted as input for the shared memory method. Simultaneously, the distributed memory method must maintain global identifiers specifying duplicate data between subdomains. Duplicate data are specified using global identifiers similar to the approach in \cite{NAVE2004191}. These global identifiers are essential when receiving and scattering interface data, as they are used to identify face-connections between subdomains. Every vertex in the grid is assigned a subdomain id (during decomposition) and an integer id local to that particular subdomain. Decomposition and this assignment of data occurs on a single node, and subdomains are not packed/sent to other nodes of the HPC cluster until all data assignment is completed. During assignment, whichever subdomain encounters the point first is considered its owner, while others store this point as a duplicate and its global identifier (subdomain id and local id) information. Interface points remain untouched throughout adaptation. Any newly created points are assigned new global identifiers (based on the number of current local ids for that subdomain) at the end of adaptation. If these points are gathered and sent to a neighbor during an interface shift, their global identifiers are sent with them so that the neighbor will also have this information.

The distributed memory method is implemented in the C++ programming language and as such, the standard library memcpy function is utilized for the packing/unpacking of data during migration. It is preferable to organize subdomain data into the simplest data structures possible (arrays, plain old data types, etc.) to make the utilization of this function seamless; otherwise, careful attention must be given to more complex data structures, as they are more likely to induce memory errors if not handled correctly (e.g. container types, dynamically allocated pointers, etc.). These simple distributed memory method data structures are converted into the complex data structures utilized by the shared memory code (linked lists of dynamically allocated pointers to class objects representative of mesh elements, for example) when jump-starting a subdomain for adaptation, and vice-versa after a subdomain has completed adaptation (for an upcoming interface shift).

\subsection{Requirements for the Distributed Method's Design}
The below requirements serve as a guide to understanding how a shared memory code should be designed if one wishes to use it as a black box for distributed computing. These are based on lessons learned in needing to re-design the shared memory CDT3D code, and will be explored in more detail (giving specific cases) in the final manuscript.

\begin{enumerate}
    \item Operations should be designed to execute successfully based only on subdomain input data provided (without assuming access to the entire input domain).
    \item If a subdomain becomes larger or smaller throughout program execution, the accompanying data structure(s) utilized within the shared memory code must also be updated before any subsequent processing of that subdomain, i.e., MII adaptation.
    \item All operations should be implemented using the speculative execution model (locking/unlocking of data), thus allowing the precursory freezing of specific data without needing to further modify each operation to process a new input type (interface data).
    \item Point creation/insertion on elements adjacent to interface elements inhibits grid generation convergence; therefore, a buffer zone must be established around locked (interface) elements to reach convergence.
    \item Assumptions regarding operation input must be identified and successful processing should remain consistent regardless of input size.
    \item The order in which operations are designed to be executed in the shared memory method should also be reflected in their order of execution in the distributed method.
\end{enumerate}

\section{Preliminary Results}
\subsection{Grid Adaptation Method}
The same grid adaptation method utilized in \cite{TsolakisEvaluation}, for the shared memory CDT3D's evaluation and comparison to other parallel meshing strategies, is utilized here for a progress and viability evaluation of the distributed memory method. The goal is to adapt a given grid so that it conforms to an anisotropic metric field \emph{M}. A comprehensive introduction to the definition and properties of the metric tensor field is provided in \cite{LoseilleAlauzetCMFI}. The complexity \emph{C} of a continuous metric field \begin{math}\begin{mathcal}M\end{mathcal}\end{math} is defined as:

\begin{equation}
C(\begin{mathcal}M\end{mathcal}) = \int_\Omega \sqrt{det(\begin{mathcal}M\end{mathcal}(x))}dx.
\end{equation}

Complexity on the discrete grid is computed by sampling \(\begin{mathcal}M\end{mathcal}\) at each vertex \emph{i} as the discrete metric field \emph{M},

\begin{equation}
C(M) = \sum_{i=1}^{N} \sqrt{det(M_i)}V_i,
\end{equation}

where \(V_i\) is the volume of the Voronoi dual surrounding each node. The complexity of a grid is known to have a linear dependency with respect to the number of points and tetrahedra, shown theoretically in \cite{LoseilleAlauzetCMFI} and experimentally verified in \cite{LoseilleAlauzetCMFII}\cite{Park2015ComparingAO}. The number of vertices are approximately 2\emph{C} while the number of tetrahedra are approximately 12\emph{C}. As shown in \cite{LoseilleAlauzetCMFI}\cite{TsolakisEvaluation}, scaling the complexity of a metric can generate the same relative distribution of element density and shape over a uniformly refined grid compared to the original complexity. The metric tensor \(M_{C_r}\) that corresponds to the target complexity \(C_r\) is evaluated by \cite{LoseilleAlauzetCMFI}:

\begin{equation}
    M_{C_r} = \left(\frac{C_r}{C(M)}\right)^\frac{2}{3} M,
\end{equation}

where \emph{M} is the metric tensor before scaling and C(M) is the complexity of the discrete metric before scaling.

In order to evaluate the progress made in the distributed memory method's implementation, quantitative results are examined with respect to overhead incurred by the interface shift operation, as opposed to alternative methods which use global synchronization and re-partitioning to process interface elements during mesh generation. Qualitative results are examined with respect to metric conformity of the adapted mesh. These qualitative measures described below are the same as those used by the Unstructured Grid Adaptation Working Group\footnote{\url{https://ugawg.github.io/}}. The adapted meshes of the early distributed memory method are compared to those of the shared memory method in order to verify the viability of performing interface shifts in the distributed memory method and its impact on the method's stability. 

The aim of metric conformity is the creation of a unit grid, where edges are unit-length and elements are unit-volume with respect to the target metric. For calculating edge length, we adopted the same definition that appears in \cite{ALAUZETSizeGradation}. For two vertices \emph{a} and \emph{b}, an edge length in the metric \(L_e\) can be evaluated using:

\begin{equation}
\begin{split}
& L_e = 
\begin{cases}
    \frac{L_a - L_b}{log(L_a/L_b)} & |L_a - L_b| > 0.001 \\
    \frac{L_a+L_b}{2} & otherwise
\end{cases} \\
& L_a = (v_e^TM_av_e)^{\frac{1}{2}},L_b = (v_e^TM_bv_e)^{\frac{1}{2}}
\end{split}
\end{equation}

and an element mean ratio shape measure can be approximated in the discrete metric as:

\begin{equation}
    Q_k = \frac{36}{3^{1/3}} \frac{\left(|k|\sqrt{det(M_{mean})}\right)^\frac{2}{3}}{\sum_{e\epsilon L} v_e^TM_{mean}v_e},
\end{equation}

where \emph{v} is a vertex of element \emph{k} and \(M_{mean}\) is the interpolated metric tensor evaluated at the centroid of element \emph{k}. Since the goal is to create edges that are unit-length, edges with length above or below one are considered to be sub-optimal. The measure for mean ratio is bounded between zero and one since it is normalized by the volume of an equilateral element. One is the optimal quality for an element's mean ratio shape.

\subsection{Experimental Setup}
PREMA 2.0, shared memory CDT3D, and the early distributed memory CDT3D codes were all compiled using the GNU GCC 7.5.0 and Open MPI 3.1.4 compilers. Data were collected on Old Dominion University's Wahab cluster using dual socket nodes that each featured two Intel\textsuperscript{\textregistered} Xeon\textsuperscript{\textregistered} Gold 6148 CPUs @ 2.40 GHz (20 slots) and 384GB of memory.

\subsection{Delta Wing Geometry}
\begin{figure}[htb]
\begin{center}
\includegraphics[width=0.65\textwidth]{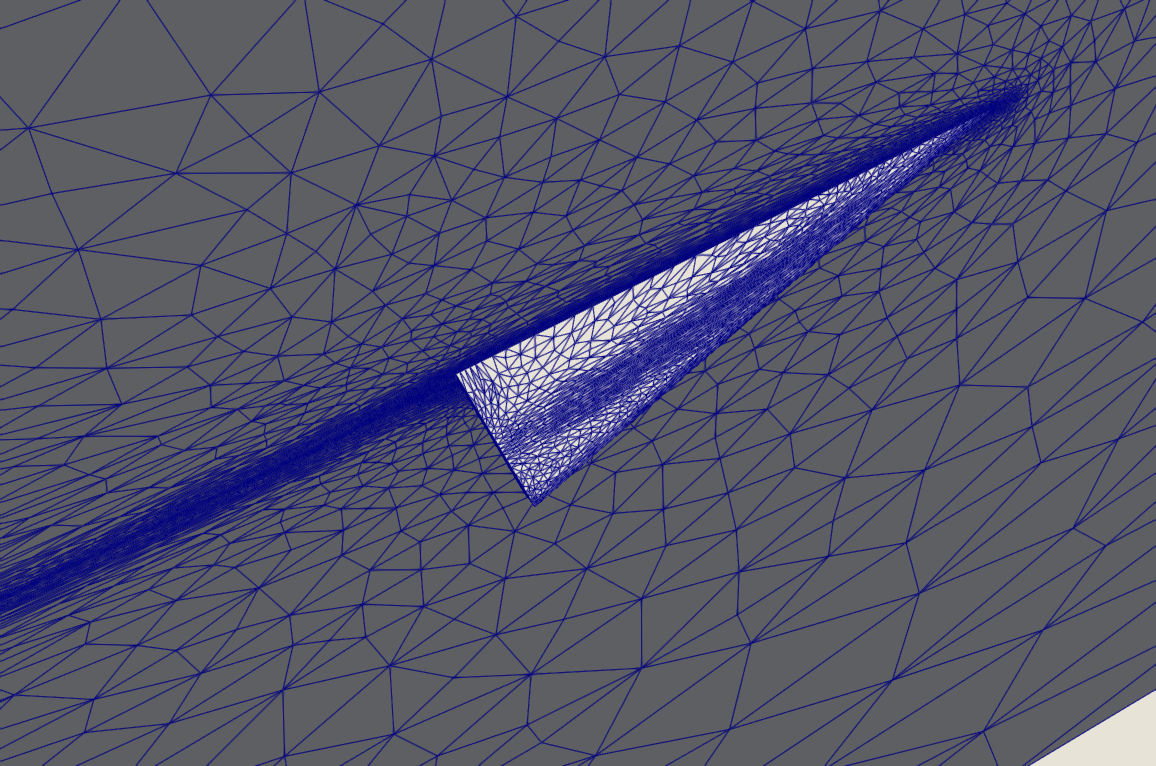}
\end{center}
\caption{Delta wing with multiscale metric derived from a laminar flow and a complexity of 50,000}
\label{fig:delta_wing}
\end{figure}

Figure \ref{fig:delta_wing} shows a delta wing geometry with a solution-based metric field derived from a laminar flow. This geometry was also utilized in the aforementioned shared memory CDT3D evaluation study \cite{TsolakisEvaluation}. 

The delta wing geometry is made of planar facets. Its multiscale metric \cite{ALAUZET2010561} is constructed based on the Mach field of a subsonic laminar flow. The initial grid is adapted to a complexity of 50,000 and is scaled to a complexity of 5,000,000 (generating approximately 50 million elements) for a quantitative evaluation. It is scaled to a complexity of 500,000 (generating approximately 5 million elements) for a qualitative evaluation (for reasons discussed below). Details of the verification of the delta wing/grid adaptation process is provided in \cite{VerificationUGAComponents}.


\subsubsection{Qualitative Improvement from Interface Shifts and MII Adaptation}
\begin{figure}[htb]
     \centering
     \begin{subfigure}[htb]{0.45\textwidth}
         \centering
         \includegraphics[width=\textwidth]{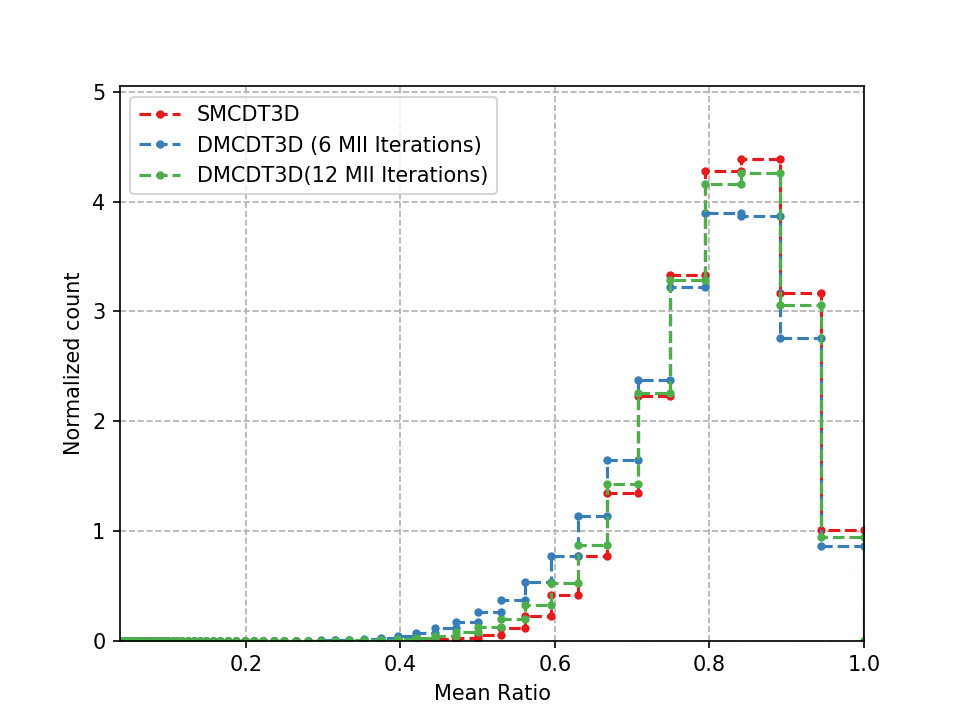}
     \end{subfigure}
     \hfill
     \begin{subfigure}[htb]{0.45\textwidth}
         \centering
    \includegraphics[width=\textwidth]{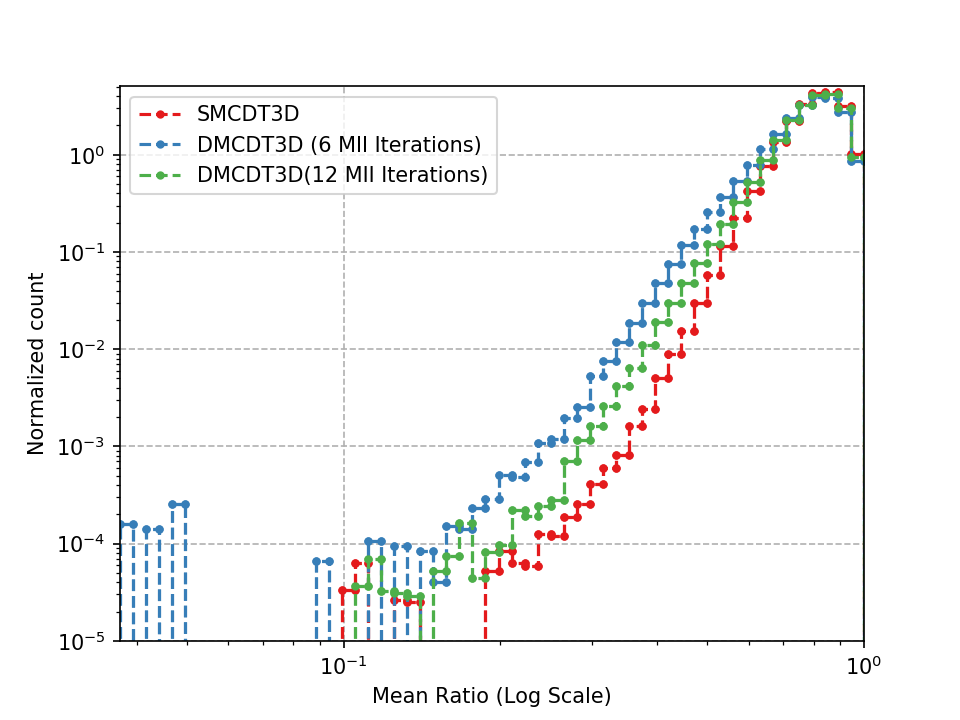}
     \end{subfigure}
        \caption{Comparison of the mean ratio of elements within the generated grids, utilizing a decomposition of 128 subdomains, for the delta wing at 500,000 complexity in linear and logarithmic scales}
        \label{fig:MR}
\end{figure}

\begin{figure}[htb]
     \centering
     \begin{subfigure}[htb]{0.45\textwidth}
         \centering
         \includegraphics[width=\textwidth]{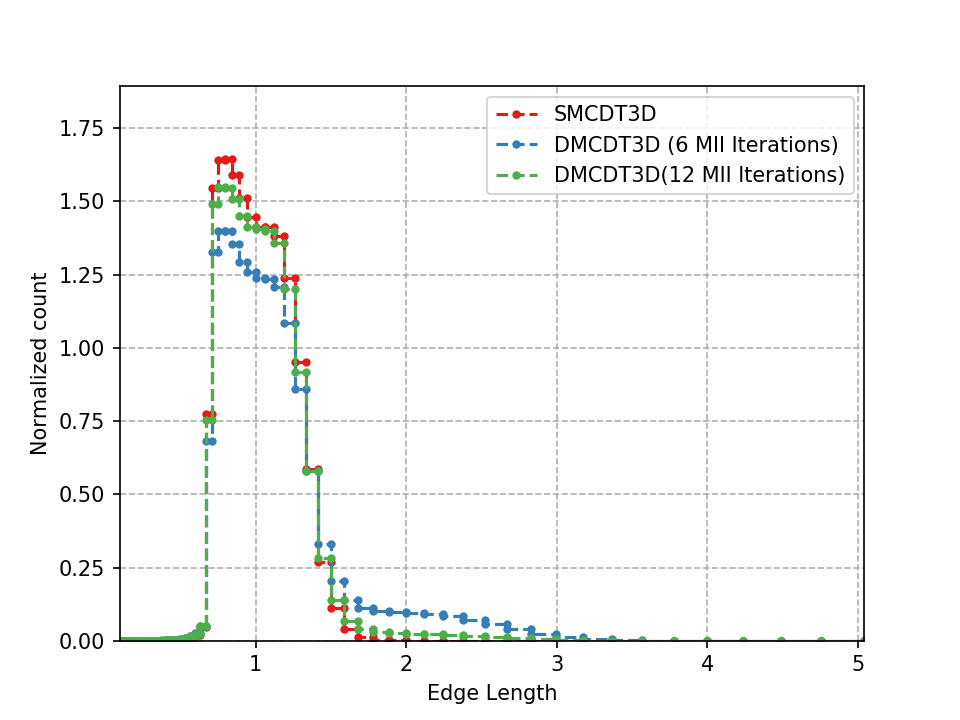}
     \end{subfigure}
     \hfill
     \begin{subfigure}[htb]{0.45\textwidth}
         \centering
    \includegraphics[width=\textwidth]{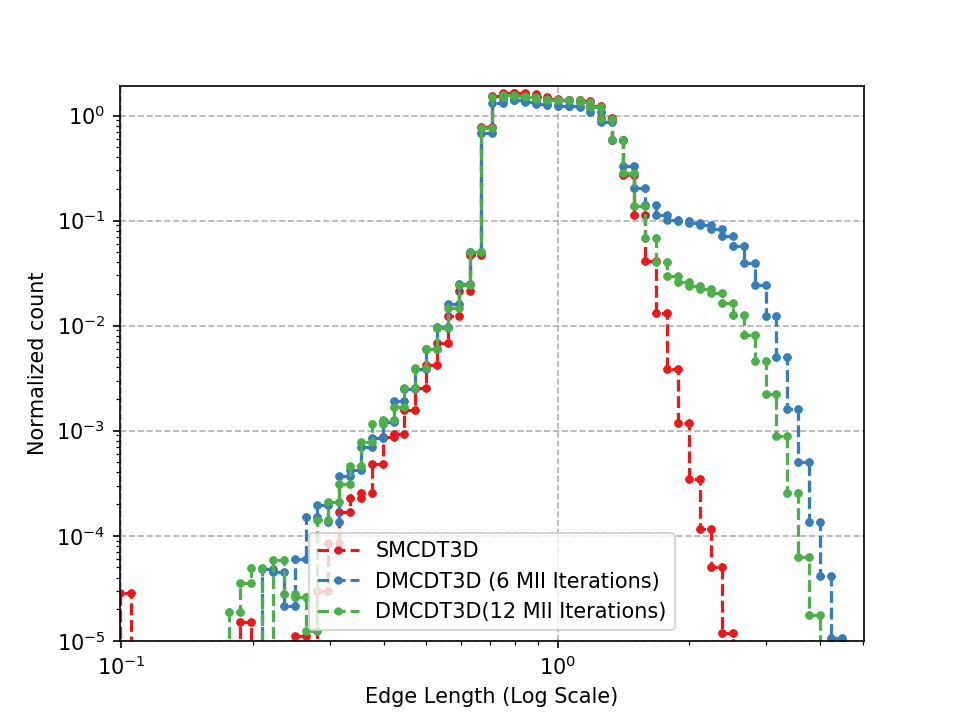}
     \end{subfigure}
        \caption{Comparison of the edge lengths of elements within the generated grids, utilizing a decomposition of 128 subdomains, for the delta wing at 500,000 complexity in linear and logarithmic scales}
        \label{fig:EL}
\end{figure}

There are several factors that affect the distributed memory method's output grid's metric conformity and scalability - the method of decomposition, the number of interface shift iterations, and the number of layers of elements gathered to use for the aforementioned "buffer zone" surrounding cavities needed for interface element adaptation. The optimal settings of these heuristics are undergoing investigation, as they vary between different geometries and require careful consideration when meshing larger geometries. Consequently, qualitative results are shown for the delta wing geometry adapted at 500,000 complexity. Qualitative results adapted at the larger complexity will be provided in the final manuscript. 
Metric conformity, characterized by element shape measure and edge length histograms, for the output meshes generated by the shared memory CDT3D and distributed memory method are compared in figures \ref{fig:MR} and \ref{fig:EL}. The results from the distributed memory method were generated from a decomposition (strictly across the x-axis using PQR) of 128 subdomains and 5 layers of elements gathered by subdomains that have not undergone MII adaptation (and 10 layers gathered by those that have). Subdomains that have undergone MII adaptation will need to send more layers in order to include their newly adapted elements; otherwise, they are likely to simply send back the non-adapted layers they had received in a previous iteration. Each graph also shows the distributed memory method's results after performing 6 MII adaptation iterations vs. 12 iterations. When performing 12 MII adaptation iterations, the grid generated by DMCDT3D exhibits good overall quality similar to that generated by the original shared memory method. 

\subsubsection{Relative Performance When Avoiding Global Synchronization}
As stated previously, the distributed memory method aims to avoid collective communication. PREMA gives the method the capability to avoid any explicit or implicit global synchronization. Establishing dependencies between mobile objects (subdomains) helps facilitate the execution of interface shifting using PREMA's event system. As soon as all relevant subdomains (within a neighborhood) have completed interior adaptation, they are colored to distinguish which will send or receive interface data, the shift is performed between corresponding neighbors, colored subdomains are adapted, and the topology of neighboring subdomains is updated. Any processing element, with regards to a particular subdomain, has the flexibility to proceed in its execution of these steps as long as the dependencies of this subdomain are satisfied (i.e. are available for communication whether it's an interface shift or topology update). While the current implementation of the distributed memory method utilizes a centralized model in testing the correctness and stability of the method (e.g. the master process colors subdomains after workers have communicated amongst themselves and finally the master about the topology update), its final implementation will utilize a decentralized variant of Luby's algorithm to create maximal independent sets of subdomains \cite{Lubysalgorithm}, thus allowing neighborhoods of subdomains to operate independently of each other (while working in tandem with PREMA's event functionality) without needing to communicate with the master process for coloring.

\begin{figure}[htb]
\begin{center}
\includegraphics[width=0.7\textwidth]{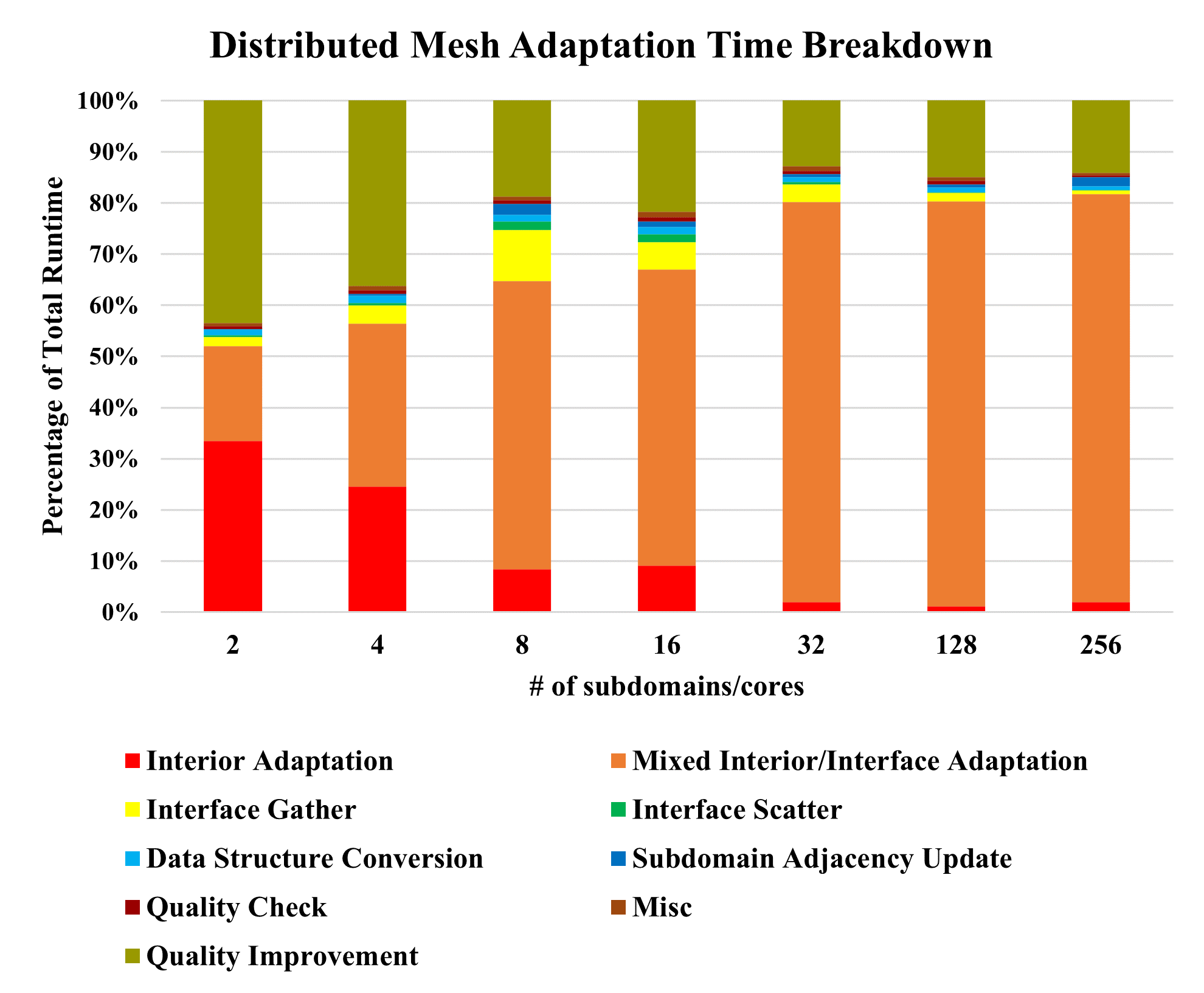}
\end{center}
\caption{Percentage Breakdown of Adaptation Time for the Distributed Memory Method in Adapting Delta Wing Geometry at 5,000,000 complexity}
\label{fig:percentage_adaptation_breakdown}
\end{figure}

Figure \ref{fig:percentage_adaptation_breakdown} shows the relative performance costs of communication within the distributed memory method when adapting the delta wing geometry at 5,000,000 complexity, with regards to executing 6 iterations of interface shifts and MII adaptation. The communication surrounding the adaptation of mixed interior/interface data includes: coloring subdomains, gathering interface data (for those subdomains designated to send), scattering interface data (on those subdomains designated to receive), updating the topology of subdomain adjacency, converting between the distributed memory data structures and the shared memory method's data structures, checking the quality of subdomains (to influence coloring), and miscellaneous operations (declaring variables, resizing data structures, cleanup, profiling, etc.). In each case presented in figure \ref{fig:percentage_adaptation_breakdown}, the number of subdomains matches the number of processing elements (e.g. 16 subdomains/16 cores, 256 subdomains/256 cores, etc.). The percentages represent the sum runtime spent performing each particular operation. While these operations occupy more time on lower configurations of subdomains/cores, they become smaller as the number of subdomains/cores are increased, occupying less than 15\% of total runtime on average and approximately 5\% of total runtime on larger configurations of cores. As opposed to traditional communication methods that induce overhead of more than 50\% (as reported in the ECP study \cite{ECPMPIStudy}), the cost of communication in permitting the adaptation of interface elements in the distributed memory method saves approximately 35\% - 45\% of communication overhead thus far.

The percentage of runtime occupied by mixed interior/interface adaptation occupies much of the adaptation time due to the fact that a subdomain becomes larger after scattering interface data and therefore requires more processing time in adaptation. This is an ongoing challenge that we intend to address in the future. The shared memory software requires significant time when processing elements that have already been adapted, resulting in longer adaptation times for subdomains than the original interior adaptation times. The shared memory software will likely undergo some further redesign to remedy this behavior (and the ensuing lesson from these changes will be included as another requirement for the distributed design). The requirement regarding the order of operations stems from this issue. The shared memory CDT3D method utilizes several mesh operations to accomplish adaptation. Understanding the nature of these operations and their intended use becomes vital for the distributed method. Originally, all operations (in figure \ref{fig:cdt3d-anisotropic-pipeline}) were executed over each subdomain within each iteration of adaptation. When performing MII adaptation, worst performance than our current results would occur. This is because the edge collapse operation is designed to be a pre-refinement and post-refinement operation. It initially overcoarsens a domain (because this can potentially lead to better end quality for the mesh) and in the end removes short edges created during refinement/adaptation. Executing this operation repeatedly over the same sets of data (shifted with interface data between subdomains) causes the shared memory method to repeatedly overcoarsen and then recreate/re-insert new points over previously adapted elements that already satisfied qualitative criteria. Therefore, the distributed memory method now utilizes the beginning edge collapse operation only when performing the initial interior adaptation pass. It does not utilize a final edge collapse operation until the interface shifting phase has completed. After this final edge collapse, the quality improvement phase commences over subdomains. Preserving this order of operations improved the MII adaptation runtime but has not fully remedied the problem. This is why quantitative data regarding scalability is not yet shown. It must also be noted that the percentages shown in figure \ref{fig:percentage_adaptation_breakdown} are relative to the current implementation. Once this particular problem has been remedied, all other percentages of operations are likely to increase. However, it should also be noted that the communication operations will be further optimized in the final implementation.

Sequential pre-processing time was a consistently small fraction across all runs, as data decomposition maintained a runtime of approximately 1 second. Time spent coloring subdomains was also a consistently small fraction of total runtime (less than 1 second across all runs). While producing meshes of sufficient quality over multiple interface shifts and MII adaptation, more detailed statistics (timing profiles of operations across different configurations of cores) will be provided in the final manuscript.



\section{Conclusion}
The foundational elements of a distributed memory method for mesh generation are presented. Meshing functionality is separated from performance aspects in order to fit a scalable framework that is designed to leverage maximum concurrency offered by large-scale architectures. Mesh adaptation is handled by a shared memory code called CDT3D. Several requirements regarding the distributed method's design are given based on lessons learned from redesigning the shared memory code, enabling its integration into the distributed memory method. Most importantly, all major operations within the shared memory code are designed to adopt the speculative execution model, enabling the strict adaptation of interior subdomain elements so that interface elements can be adapted in a separate step to maintain mesh conformity. Interface elements undergo several iterations of shifting so that they are adapted when their data dependencies are resolved. Communication, work load balancing, and migration of data are handled by a parallel runtime system called PREMA. PREMA offers a particularly useful "event" feature which aids in establishing data dependencies between subdomains, thus enabling "neighborhoods" of subdomains to work independently of each other in performing interface shifts and adaptation.

Preliminary results show that after several passes of interface shifts and adaptation, the distributed memory CDT3D method is able to produce meshes of comparable quality to those generated by the original shared memory CDT3D code. Relative communication performance also suggest that the interface shift operation presents a potentially viable solution in achieving scalability for mesh adaptation when targeting configurations with large numbers of cores (building upon the shared memory CDT3D method which only utilized up to 40 cores in an earlier study \cite{TsolakisEvaluation}). Given the costly overhead of collective communication identified within the study of DoE's Exascale Computing Project \cite{ECPMPIStudy, Klenk2017ECPStudy} and seen in existing state-of-the-art software \cite{TsolakisEvaluation}, the distributed memory method's emphasis on avoiding global synchronization will likely prove beneficial (and shall be further tested on larger configurations of cores with larger geometries in the finalized manuscript of this work, providing more comprehensive quantitative data).

\section*{Acknowledgments}
This research was sponsored by the Richard T. Cheng Endowment, the Southern Regional Education Board (SREB) State Doctoral Scholar Fellowship, and the National Institute of General Medical Sciences of the National Institutes of Health under Award Number 1T32GM140911-03. The content is solely the authors’ responsibility and does not necessarily represent the official views of the National Institutes of Health.

\bibliography{references}

\end{document}